\documentclass{elsart}
\usepackage{latexsym}
\usepackage{epsfig}
\usepackage{natbib}

\def\url#1{#1} 

\begin{document}

\begin{frontmatter}

\title{The relation between far-UV and visible extinctions.}

\author{Fr\'ed\'eric \snm Zagury}
\address{
\cty 02210 Saint R\'emy Blanzy, \cny France
\thanksref{email} }

   \thanks[email]{E-mail: fzagury@wanadoo.fr}


 \begin{abstract}
For directions of sufficient reddening ($E(B-V)>\sim 0.25$), 
there is a simple relation between the slope of the extinction curve 
in the far-UV and $E(B-V)$.
Regardless of direction, the far-UV extinction curve is 
proportional to $1/\lambda^ne^{-2E(B-V)/\lambda}$ ($\lambda$ in 
$\mu$m, $n=4$), in accordance with 
the idea that reddened stars spectra are 
contaminated by scattered light \citep{uv4}.

This relation is not compatible with the standard theory of 
extinction which states that far-UV and visible extinctions are due 
to different classes of grains.
In that model the two (far-UV and visible) extinctions vary thus 
independently according to the proportion of each type of grains.

In preceding papers I have shown that the standard theory
cannot explain UV observations of nebulae, and is contradicted by the UV spectra of stars 
with very low reddening:
for how long shall the standard theory be considered as the 
interpretation of the extinction curve? 

\end{abstract} 
 \begin{keyword}
{ISM: Dust, extinction}
     \PACS
     98.38.j \sep
98.38.Cp\sep
98.58.Ca
  \end{keyword}   
\end{frontmatter}
 \section{Introduction} \label{intro}
Extinction curves normalized by $E(B-V)$ are similar in the 
visible part of the spectrum, but they exhibit noticeable differences in the 
far-UV \citep{savage85}.

If the light we receive from a reddened star is only direct 
starlight, as it is supposed by the standard theory of extinction,
 the optical and the far-UV extinctions must be caused by different types of grains.
The important variations observed in the far-UV slope of the normalized 
extinction curves are due to the different proportions of each type of 
grain from one region to another.

I have shown in a previous series of papers the 
possibility for the far-UV light that we receive from the direction of 
a reddened star, to be 
mainly light scattered at small angular distances from the star and re-injected into the beam of 
the observation.
If scattering is due to particles small enough to have equal 
reflectance, then the scattering is coherent within a very small cone 
centered on the star, and the intensity of the scattered 
light is -unless multiple scattering is important- proportional to the square of the number of 
particles within the cone.

One important consequence of this interpretation is that 
the linear visible extinction of starlight extends to the UV.
This was verified up to the bump region for stars with 
not too high reddening ($E(B-V)\le 0.5$, \citet{uv2}), for which 
scattered light is important in the far-UV only.

In the far-UV, scattering is enhanced because extinction, and the 
number of photons available for scattering, is higher.
Coherent scattering considerably amplifies the intensity of the 
scattered light, which - assuming a linear extinction 
law in the far-UV- can reach around $20\,\% $ of the direct starlight corrected for extinction \citep{uv2}.
This is by far larger than the reddened direct starlight if extinction 
is important.
For example, $E(B-V)\sim 0.5$ yields a ratio at $\lambda=1500\,\rm\AA$ of scattered light to 
direct starlight of $\sim 0.2/e^{-2E(B-V)/\lambda}=160$ ($\lambda$ in 
$\mu$m, see \citet{uv2}).
Of course, if extinction is large, either because $E(B-V)$ is 
large or towards the shortest wavelengths, scattered light is finally
extinguished, as direct starlight is.

For large enough $E(B-V)$ 
the far-UV light we receive from a star  is nearly all scattered light.
If scattered light is extinguished like direct starlight as 
$e^{-2E(B-V)/\lambda}$, and
since the scattering cross section of the small particles varies as 
$1/\lambda^n$ ($n\ge 4$, \citet{jackson}), we expect the spectrum of the 
scattered light -and the far-UV spectrum of a star with sufficient reddening-
to vary as $\lambda^{-n}e^{-2E(B-V)/\lambda}$ \citep{uv3}.
We see that the slope of the far-UV extinction curve is now related to the slope 
of its' optical part.

In this framework, the reduced spectrum of a reddened star (the spectrum of the star divided by the spectrum of 
an unreddened star of same spectral type, \citet{uv2})  multiplied by 
$\lambda^n$ must be an exponential; the exponent determines the 
far-UV slope of the reduced spectrum, independently of the exact 
value of $n$.
$n=4$, the Rayleigh scattering case, 
gave a good fit to the spectrum of HD46223 \citep{uv3} and is adopted 
hereafter.

The idea of this paper is to take stars with sufficient reddening, 
and to study the dependence on $\lambda$ of their far-UV reduced spectrum multiplied by 
$\lambda^4$.
The product is expected to decrease exponentially.
Determination of the exponent $E_{uv}$ gives an estimate of $E(B-V)$.
$E_{uv}$ can be compared to the value, $E_{vis}$, found for $E(B-V)$ from the 
spectral type of the star and its' optical color $B-V$. 
I expect $E_{uv}$ and $E_{vis}$ to have close values.
Possible contamination of direct starlight by scattered light in the visible
will give an observed color of a star which underestimates the exact 
$B-V$ \citep{uv3}.
It implies that $E_{vis}$ can be slightly smaller than $E_{uv}$ \citep{uv3}.
The exact value of $E(B-V)$ is given by $E_{uv}$.

Section~\ref{data} presents the data.
The determination of $E(B-V)$ from the far-UV spectra and the comparison with 
the values derived from the optical properties of the stars is made 
in section~\ref{comp}.
The result is discussed in section~\ref{dis}.
A summary is given in the conclusion.
\section{Data} \label{data}
The stars of Table~\ref{tbl:stars} correspond to a wide range of 
different extinction curves and reddenings.
These stars belong to existing atlases of extinction curves 
\citep{savage85, fitz90, aiello} that I have used in the past three years. 
Stars with too low reddening 
($E(B-V)< \sim 0.2-0.3$) are not suitable to the purpose of the paper: 
for these stars direct 
starlight is not negligible in the UV and 
the reddening of the comparison stars becomes an important source of 
uncertainty.
They have been discarded.

Some particular objects, like the Red Rectangle, were not included.
The current comprehension of the Red Rectangle \citep{waelkens}, 
is that HD44179 is hidden behind a thick disk of interstellar matter; 
we do not receive direct starlight from the 
direction of the star but only light scattered from the poles of the 
disc.

The far-UV spectra were retrieved from the IUE database 
(\url{http://iuearc.vilspa.esa.es}).
For each star an unreddened star of close spectral type was found and 
used to obtain the reduced spectra.
The numbers in column~7 of Table~\ref{tbl:stars} refer to the 
comparison star number of Table~\ref{tbl:ref}.

$B-V$ comes from SIMBAD database (\url{http://simbad.u-strasbg.fr}). 
In general these values agree with Tycho satellite estimates (also 
available at SIMBAD), after multiplication by $\sim 0.92$  
to take into account the difference of Tycho and Johnson filters.
\section{Data analysis} \label{comp}
\begin{figure*}[p]
\resizebox{\hsize}{!}{\includegraphics{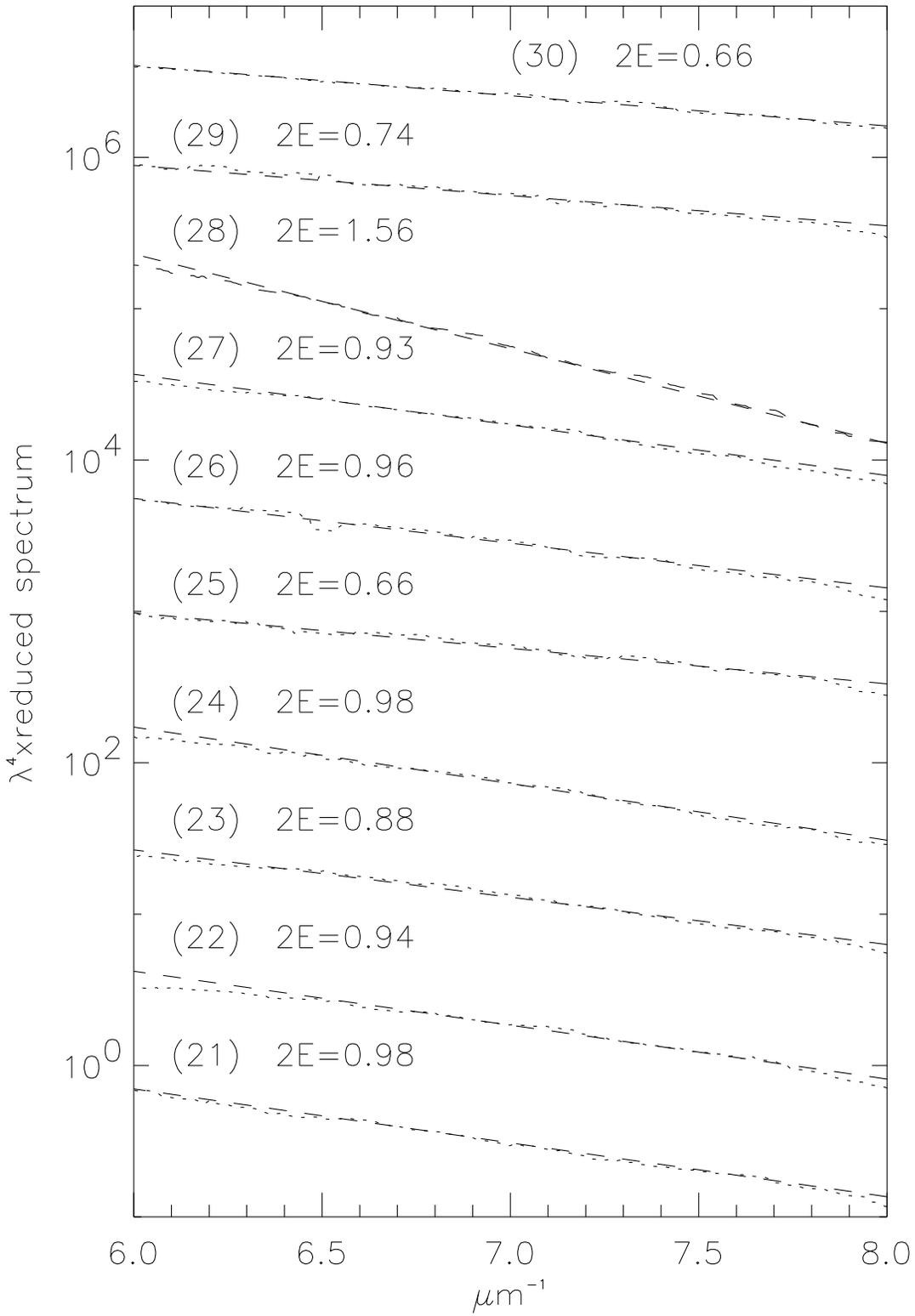}} 
\caption{Far-UV reduced spectra of the reddened 
stars multiplied by $\lambda^4$. 
The number above each curve corresponds to the star number, first column of 
Table~\ref{tbl:stars}.} 
\label{fig1}
\end{figure*}
\begin{figure*}[p]
\resizebox{\hsize}{!}{\includegraphics{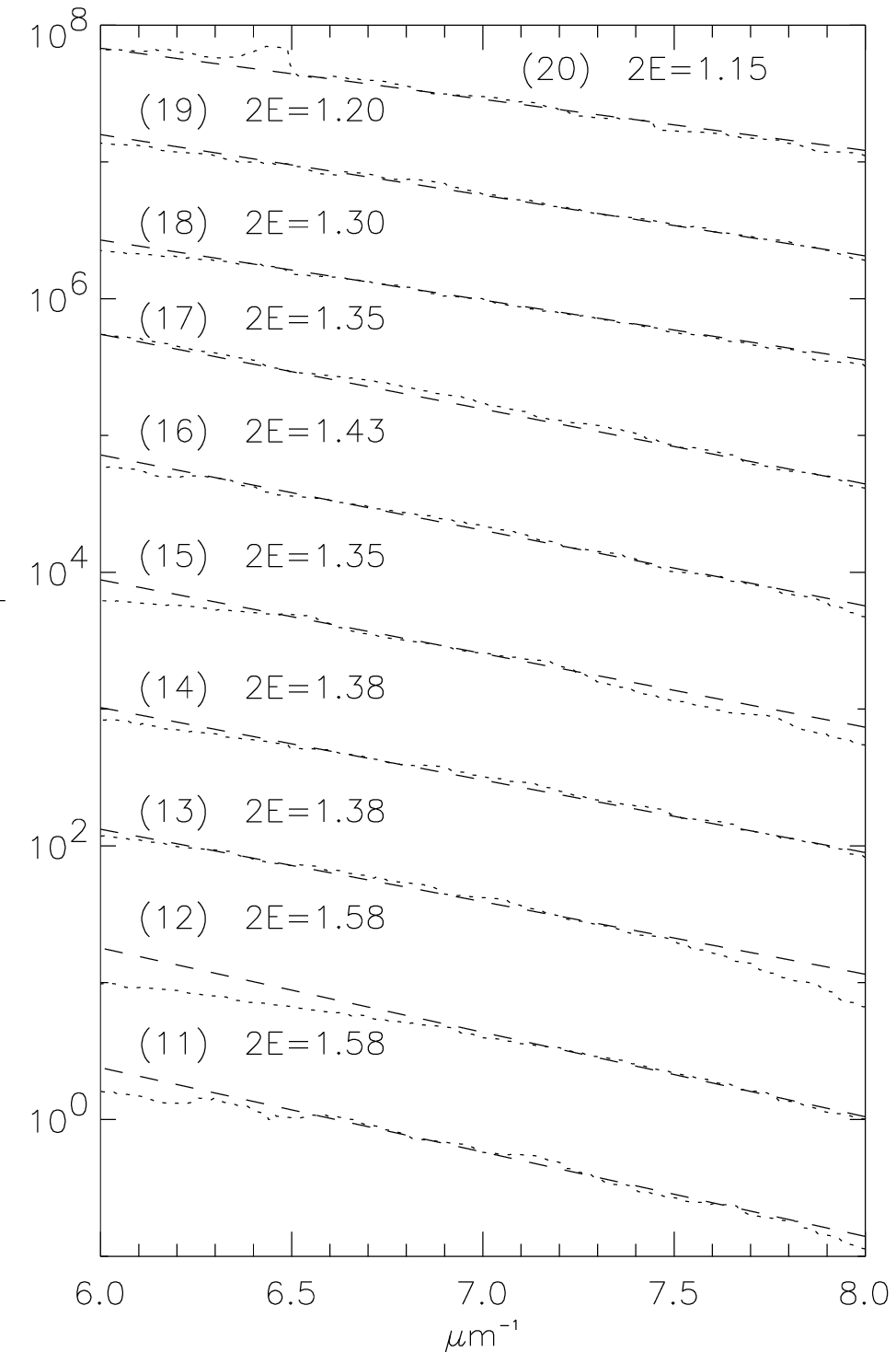}} 
\caption{Figure~\ref{fig1} continued. } 
\label{fig2}
\end{figure*}
\begin{figure*}[p]
\resizebox{\hsize}{!}{\includegraphics{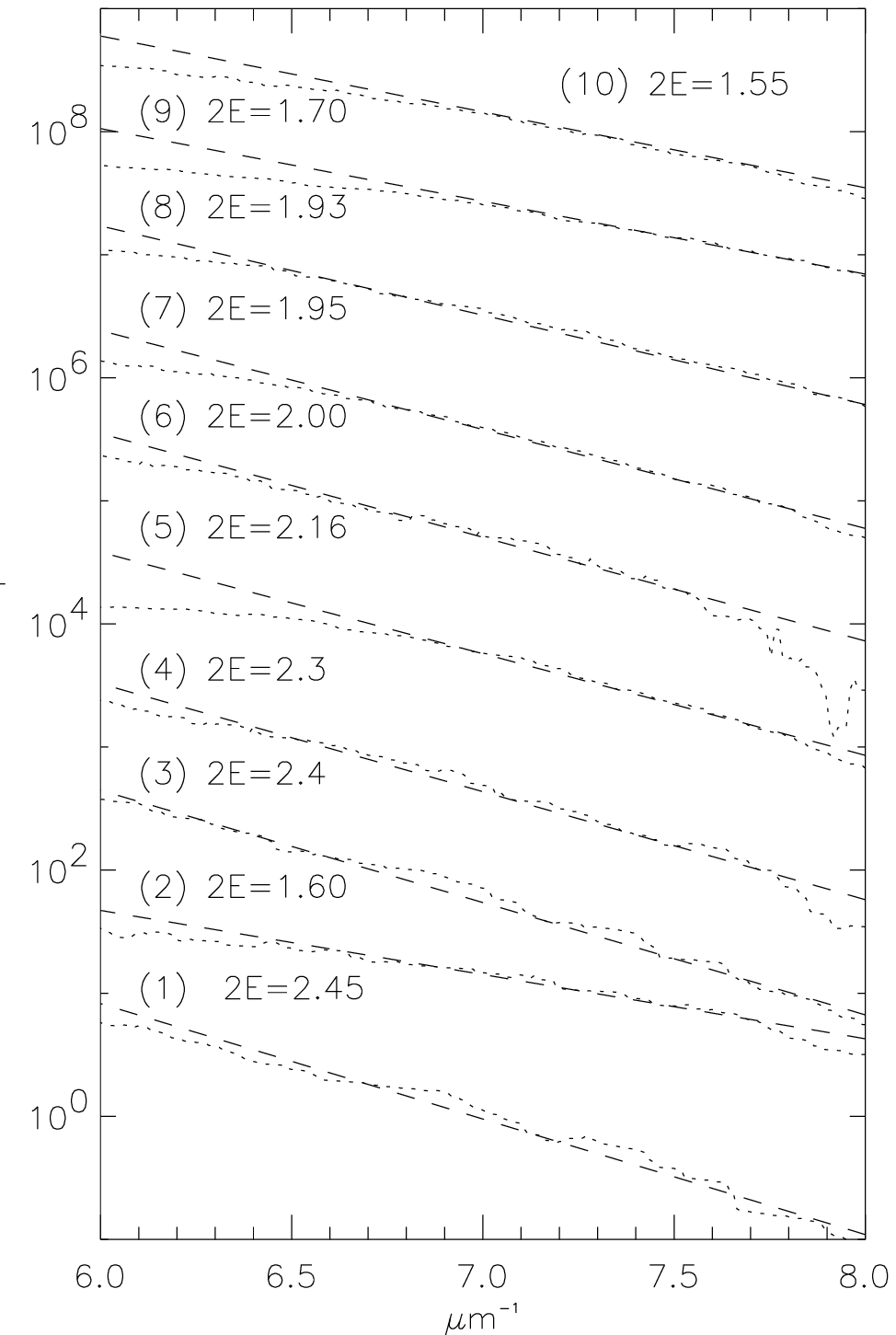}} 
\caption{Figure~\ref{fig1} continued.} 
\label{fig3}
\end{figure*}
\begin{figure*}
\resizebox{!}{0.8\columnwidth}{\includegraphics{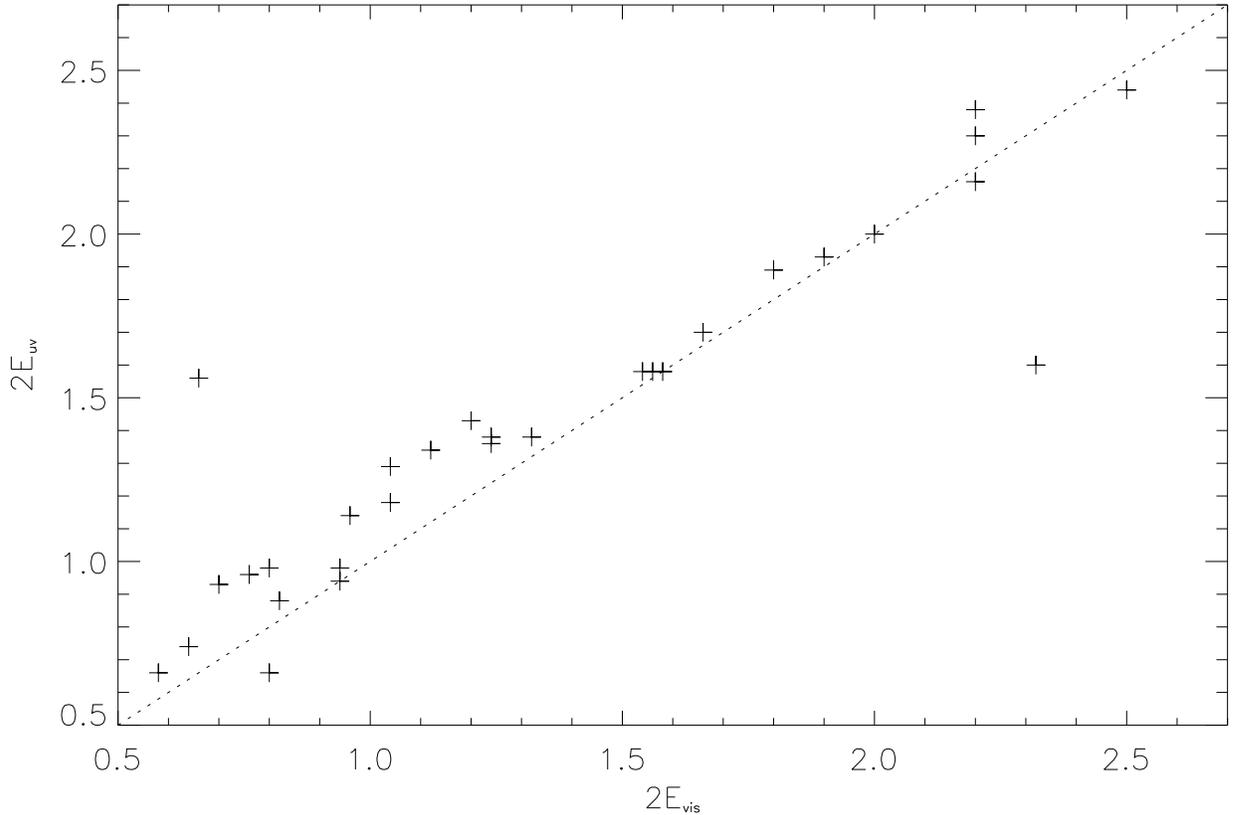}} 
\caption{Plot of $2E_{uv}$ versus $2E_{vis}$.
} 
\label{fig4}
\end{figure*}
The far-UV reduced spectra, multiplied by $\lambda^4$ are plotted,
figure~\ref{fig1} to figure~\ref{fig3}, with an 
arbitrary scaling factor along the y-axis (logarithm y-coordinate).
Comparison stars have been corrected for the 
extinction in their direction, as indicated in \citet{uv5}.
This correction is e.g. smaller or of the same order than the uncertainty on $E_{uv}$ and 
cannot affect the results of the paper.

The resulting spectra decrease exponentially in the far-UV 
(figures~\ref{fig1} to \ref{fig3}).
The exponent $2E_{uv}$ is written 
after the number of the star and reported in the sixth column of Table~\ref{tbl:stars}.

Figure~\ref{fig4} plots $2E_{uv}$ versus $2E_{vis}$.
This figure has three main sources of uncertainty.

A first source of error is a possible mismatch on the spectral type 
of the stars, and/or  a difference between the spectral types of the 
star and of the comparison star. 
Except for an unprobable large mistake on the spectral type, the value of $2E_{vis}$ will not be 
alterated too much.
But, spectral type mismatch can noticeably change the slope $2E_{uv}$ of the reduced 
spectrum of the star. 
For \citet{cardelli88} HD62542 is a B5V .
If the star is a B3V \citep{houk78}, the cross representing HD62542 on 
figure~\ref{fig4} will be shifted up by 0.3~mag..
These errors are discussed in several papers \citep{bless72, nandy75}.
They can represent a few 0.1~mag., but they are difficult to quantify.
They are not included in the error margin given for $2E_{uv}$ (Table~\ref{tbl:stars}).

The error on $2E_{uv}$ given in Table~\ref{tbl:stars} corresponds 
to the uncertainty once a spectral type is adopted.
For the stars with the largest reddenings (figure~\ref{fig3}), the 
increasing importance of the $2200\,\rm\AA$ bump on the long 
wavelength side, and the Ly$\alpha$ emission on the short wavelength side of the spectrum, 
render difficult a precise determination of the slope.
Consequently, the uncertainty on $2E_{uv}$ is increased.

Last, some objects are special.
For the Red Rectangle (discarded here, section~\ref{data}), 
$E(B-V)$ determined from the visible magnitudes has no meaning at all, 
because the star is hidden behind a huge amount of interstellar 
matter and is not directly observed.
Particular objects should in general be stars embedded in dust clouds.

Although these sources of error can occasionally modify 
figure~\ref{fig1} to figure~\ref{fig4}, the general conclusions I 
would derive from the examination of these figures are:
\begin{itemize}
    \item  Between the $2200\,\rm\AA$ bump and the Ly$\alpha$ region the 
    (far-UV reduced spectrum)x$\lambda^4$ product can be assimilated 
    to an exponential.
   
    \item $E_{uv}$ and $E_{vis}$ are correlated and have close values. 

    \item  There is a tendency for $E_{uv}$ to be larger than $E_{vis}$.
\end{itemize}¥
This is true for all stars except for HD62542 and HD229196.

Even if the wavelength interval over which $E_{uv}$ is determined is 
short -because of the $2200\,\rm\AA$ on the one hand and the 
Ly$\alpha$ line on the other hand- the strong correlation between $E_{uv}$ 
and  $E(B-V)$ (approximated by $E_{vis}$) is remarkable enough not to be due to hazard. 
It implies a link between far-UV and visible extinctions. 

On IRAS infrared images (see also the Palomar images and \citet{cardelli88}), HD62542 is close to the edge of a 
bow-like structure, perhaps the front ridge of a shock wave.
Interstellar matter associated to HD62542 may be the reason for the 
position of HD62542 on figure~\ref{fig4}.
Note that, according to \citet{cardelli88}, the properties of extinction in the direction 
of HD62542 and in the direction of HD29647 are very similar, while the 
position of HD29647 on the plot is on the $x=y$ line.
\section{Discussion} \label{dis}
In the last thirty years, from the first publication of a large set of 
extinction curves by \citet{bless72} and the later determination of an average and normalized extinction 
curve by \citet{seaton79}, there has been several attempts to find a link 
between far-UV extinction and $E(B-V)$ \citep{savage85, fitz88, barbaro01}.
Some conclusions, for instance the dependence of far-UV extinction on 
galactic longitude (why on galactic 
longitude?) which \citet{witt84} have claimed to observe, are probably dubious.

Important deviations from Seaton's curve in the far-UV, obvious through 
visual inspection, are commonly observed.
The sample of stars used in this paper includes some of the directions which 
represent extreme cases of such deviations:
HD147165 ($\sigma$~Sco, Fig.1 of \citet{savage75}), 
HD204827 \citep{savage85,fitz88,barbaro01}, HD29647 
and HD62542 \citep{cardelli88, barbaro01}, HD37022 ($\theta^1$~Ori, \citet{cc88, barbaro01}), 
HD169454, HD37367 \citep{barbaro01}.

These examples have been used as proof of
independent variations of the three components of the extinction curve: 
visible and far-UV extinctions, and the $2200\,\rm\AA$ bump region. 
If, as it is admitted in the standard theory of extinction, the light we receive 
from reddened stars is direct starlight alone, with no addition of 
scattered light,  each of these components must be
due to a specific class of interstellar grains, present in all 
directions, but in different proportions.
This convenient way of solving the problem of interstellar 
extinction faces two major difficulties: there is up to now no 
satisfying identification of any of the three types of grains (except 
for the large grains responsible for the visible extinction, see \citet{landgraf}) and, 
there does not seem to be any logic in the grain type repartition in 
relation to environment.

It is paradoxal to see how what should be a supplementary degree of 
freedom turns to heavy constraints on the grains' composition.
Adjustment of the proportion of each type of grains permits to fit 
most individual extinction curves.
Meanwhile, the specific extinction properties each type of grains must have,
along with cosmic abundance limitations, impose 
heavy constraints on the grains' composition and on the evolution
of dust from one environment to another \citep{li97}.

The mathematical decomposition of the extinction curve proposed by \citet{fitz88} 
also fails to show a relation between far-UV and optical extinctions
\citep{fitz88, barbaro01}.
\citet{fitz88} (section~V, first paragraph) note that any attempt to relate far-UV and optical 
extinctions strongly depends upon the adopted set of fitting functions.
It is certain that within the standard theory framework and with the 
\citet{fitz88} decomposition, which for the far-UV part of the 
spectrum has no physical basis at all, there was little chance to find a 
relation between the linear optical extinction and the far-UV 
extinction.

Conclusions of section~\ref{comp} are a first step in the 
comprehension of the relation between the visible and the far-UV 
extinctions.
The relation found between the slope of the extinction curve in these two 
wavelengths domains is expected if the spectra of reddened stars are contaminated by 
scattered light and if the linear visible extinction also applies to 
UV light.
This relation exists even in the directions (mentioned in the first paragraph of this 
section) quoted to have a peculiar far-UV extinction.
\section{Conclusion} \label{conc}
I have used extinction curves in directions of sufficient 
reddening to show that far-UV and visible extinctions are related.
In these directions, extinction is strong enough for direct starlight 
to be negligible in the far-UV;
the far-UV light we receive is all scattered light.
The far-UV reduced spectrum of the stars depends on $\lambda$ as
$\lambda^{-n}e^{-2E(B-V)/\lambda}$ ($\lambda$ in $\mu$m, $n=4$).

Behind the apparent broad range of behavior so far found 
in the far-UV extinction curves \citep{bless72}, there is an underlying 
and simple order which relates far-UV extinction to visible reddening.
Accepting this relation requires to question the standard 
interpretation of the UV extinction curve.
It implies the abandonment of existing grain models, for which the 
extinction curve is the superimposition of the extinction curves of 
three separate kinds of grains.
It also questions the existence of constituents of 
interstellar matter (PAH, HAC \ldots) which these models tried to prove
using the extinction properties of interstellar dust at UV wavelengths.

The paper also gives credit to an idea I have tried to introduce in previous 
papers: the average extinction properties of interstellar dust are similar 
in most directions of space.
Deeper understanding on the truth and limits of this proposal will be 
reached through the obtention and the comparison of precise extinction 
curves covering the whole optical to far-UV domain.
{}
    \begin{table*}[p]
\caption[]{Reddened stars}		
       \[
    \begin{tabular}{|c|l|c|c|c|c|c|}
\hline
n$^{\circ}$ & name & Sp. Type & $B-V$ & $2E_{vis} \,^{\,(1)}$ & $2E_{uv} \,^{\,(2)}$ & 
Ref.${\,(3)}$\\   \hline
 1 & HD169454 & B1Ia & 0.95 & 2.50 & $2.44\pm 0.30$ & 11\\
 2 & HD229196 & O5 & 0.84 & 2.32 & $1.60\pm 0.20$ & 7\\
 3 & HD204827 & B0V & 0.80 & 2.20 & $2.38\pm 0.30$ & 10\\
 4 & BD+66 1661 & O9V & 0.81 & 2.20 & $2.30\pm 0.30$ & 4\\
 5 & HD147889 & B2III/IV & 0.84 & 2.20 & $2.16\pm 0.20$ & 8\\
 6 & HD29647 & B8III & 0.91 & 2.00 & $2.00\pm 0.30$ & 1\\
 7 & HD217061 & B1V & 0.68 & 1.90 & $1.93\pm 0.15$ & 13\\
 8 & BD+62 2125 & B1.5V & 0.65 & 1.80 & $1.89\pm 0.20$ & 11\\
 9 & HD216898 & O8.5V/9V & 0.53 & 1.66 & $1.70\pm 0.20$ & 4\\
 10 & HD217463 & B1.5Vn & 0.54 & 1.58 & $1.58\pm 0.15$ & 13\\
 11 & HD251204 & B0V & 0.48 & 1.56 & $1.58\pm 0.10$ & 13\\
 12 & BD+62 2154 & B1V & 0.51 & 1.54 & $1.58\pm 0.10$ & 13\\
 13 & HD239729 & B0V & 0.36 & 1.32 & $1.38\pm 0.20$ & 10\\
 14 & HD161056 & B1.5V & 0.37 & 1.24 & $1.38\pm 0.10$ & 13\\
 15 & HD200775 & B2Ve & 0.38 & 1.24 & $1.36\pm 0.10$ & 8\\
 16 & HD217979 & B1V & 0.35 & 1.20 & $1.43\pm 0.15$ & 13\\
 17 & HD21483 & B3III & 0.36 & 1.12 & $1.34\pm 0.20$ & 9\\
 18 & HD12993 & O6.5V/O5 & 0.20 & 1.04 & $1.29\pm 0.07$ & 11\\
 19 & HD13338 & B1V/B1III & 0.26 & 1.04 & $1.18\pm 0.10$ & 13\\
 20 & HD93250 & O5/O7 & 0.16 & 0.96 & $1.14\pm 0.10$ & 11\\
 21 & HD190944 & B1.5Vne & 0.22 & 0.94 & $0.98\pm 0.08$ & 13\\
 22 & HD147933 & B2V/B3V & 0.23 & 0.94 & $0.94\pm 0.07$ & 8\\
 23 & HD154445 & B1V & 0.15 & 0.82 & $0.88\pm 0.10$ & 13\\
 24 & HD37367 & B2IV/V & 0.16 & 0.80 & $0.98\pm 0.08$ & 8\\
 25 & HD147165 & B1III & 0.14 & 0.80 & $0.66\pm 0.05$ & 2\\
 26 & HD93222 & O7/8 & 0.07 & 0.76 & $0.96\pm 0.08$ & 11\\
 27 & HD37903 & B1.5V & 0.10 & 0.70 & $0.93\pm 0.10$ & 13\\
 28 & HD62542 & B5V & 0.17 & 0.66 & $1.56\pm 0.10$ & 3\\
 29 & HD37022 & O6pe/O7V & 0.00 & 0.64 & $0.74\pm 0.05$ & 11\\
 30 & HD38087 & B5V & 0.12 & 0.58 & $0.66\pm 0.05$ & 3\\
 \hline
 \end{tabular}    
    \]
\begin{list}{}{}
\item[$1$] $2E(B-V)$ from $B-V$ and the spectral type of 
the star.
\item[$2$] $2E(B-V)$ deduced from the UV spectrum of the star. 
\item[$3$] The number refers to the non-reddened star 
(Table~\ref{tbl:ref}) used to obtain the reduced spectrum of the star.
\end{list}
\label{tbl:stars}
\end{table*}
    \begin{table*}[p]
\caption[]{Unreddened stars}		
       \[
    \begin{tabular}{|c|c|c|c|}
\hline
n$^{\circ}$ & name & Sp. Type & $E(B-V)$ \\  \hline
 1 & HD10205 & B8III & 0.00  \\  
 2 & HD118716 & B1III & 0.03  \\  
 3 & HD199081 & B5V & 0.03  \\   
 4 & HD214680 & O9V & 0.10  \\   
 5 & HD215573 & B6IV & 0.00  \\ 
 6 & HD222661 & B9V & 0.04  \\  
 7 & HD269698 & O5e & 0.10  \\   
 8 & HD31726 & B2V & 0.03  \\   
 9 & HD32630 & B3V & 0.02  \\  
 10 & HD36512 & B0V & 0.04  \\  
 11 & HD47839 & O7V & 0.07  \\   
 12 & HD58050 & B2Ve & 0.05  \\  
 13 & HD74273 & B1.5V  & 0.04  \\ 
 \hline
\end{tabular}    
    \]
\label{tbl:ref}
\end{table*}
\end{document}